\documentclass[iop]{emulateapj}

\newcommand{\pd}[2]{\frac{\partial #1}{\partial #2}}
\renewcommand{\vec}[1]{\mathbf{#1}}

\newcommand{\DS}{\displaystyle}

\shorttitle{Numerical simulations of radiative proto-stellar jets: 
pre-ionization from X-rays}
\shortauthors{Te\c{s}ileanu et al.}

\begin{document}

   \title{Numerical simulations of radiative magnetized Herbig-Haro jets:\\ 
          the influence of pre-ionization from X-rays on emission lines}

  \author{O. Te\c{s}ileanu}
  \affil{RCAPA - Department of Physics, University of Bucharest, 
         405 Atomistilor Street, RO-077125, Bucure\c{s}ti-M\~{a}gurele, Romania}
  \affil{National Institute of Physics and Nuclear Engineering, 
         30 Reactorului Street, RO-077125, Bucure\c{s}ti-M\~{a}gurele, Romania}
  \email{ovidiutesileanu@brahms.fizica.unibuc.ro}
  \author{A. Mignone and S. Massaglia} 
  \affil{Dipartimento di Fisica Generale dell'Universit\`a di Torino, 
         Via Pietro Giuria 1, I-10125 Torino, Italy}
  \email{mignone@ph.unito.it, massaglia@ph.unito.it}         
  \and
  
  \author{F. Bacciotti}
  \affil{INAF - Osservatorio Astrofisico di Arcetri, Largo E. Fermi 5, I-50125 Firenze, Italy}
  \email{fran@arcetri.astro.it}

   \date{Received; accepted }

\begin{abstract}
We investigate supersonic, axisymmetric magnetohydrodynamic (MHD) jets with a time-dependent injection velocity
by numerical simulations with the PLUTO code. 
Using a comprehensive set of parameters, we explore different jet configurations in the attempt 
to construct models that can be directly compared to observational data of microjets.
In particular, we focus our attention on the emitting properties of traveling knots 
and construct, at the same time, accurate line intensity ratios and surface brightness maps.
Direct comparison of the resulting brightness and line intensity ratios distributions with 
observational data of microjets 
shows that a closer match can be obtained only when the jet material is pre-ionized 
to some degree. A very likely source for a pre-ionized medium is photoionization 
by X-ray flux coming from the central object.
\end{abstract}


%

\section{Introduction}
%
%
%

Jets from Young Stellar Objects (YSOs) derive their emission from the gas that has been 
heated and compressed by shocks. In fact, the actual jet matter is invisible for most of its
extension and only the cooling zones behind the shocks emit a variety of lines that can be revealed with
great accuracy and are rich of diagnostic indications on the post-shock physical parameters
such as temperature, density, ionization fraction and radial velocity.
We refer especially to the so-called ``microjets" like HH 30, DG Tau and RW Aur \citep{BE99,HM07,BA02,MA09}, 
where the line emission is limited to a region of
the jet going up to about $4^{\prime \prime}-5^{\prime\prime}$ from the forming star. Therefore, a careful
study of the shock formation and evolution is crucial to understand the physical processes at work and
to constrain jet parameters that cannot be directly observed, such as the magnetic field intensity, the
pre-shock density and temperature and the jet velocity.

Radiative shocks have been studied in steady-state conditions by several authors (e.g., \citealt{CR85}, 
\citealt{HA94}), who derived the one-dimensional post-shock behavior of various physical parameters (temperature, 
ionization fraction, electron density, etc.) as functions of the distance from the shock front. More recently,
\citet{MA05a} and \citet{TE09a} have carried out 1D numerical
studies of the time-dependent evolution of radiating, magnetized shocks. They have applied
the results to the cases of DG Tau and HH 30, with the goal to reproduce the observed behavior
of the line intensity ratios along the jet.

These studies, as discussed by \citet{Raga07} as well, brought about the problem
of the numerical resolution that is needed for a correct treatment of the post-shock region, especially
as far as the reproduction of the line ratios is concerned. To solve this problem the authors have
employed Adaptive Mesh Refinement (AMR) techniques, that allows to follow with great accuracy the
sudden temperature drop behind the shock front and save computational time.

\citet{TE09a} discussed as well the influence on the results of the cooling function details. 
They concluded that the use of a detailed treatment of radiative emissions and ionization/recombination
processes in the jet material, as well as adequate numerical resolution are very important for the 
reproduction of emission line ratios, which are extremely sensitive parameters. 
Instead, to describe the general morphology
of the jet and integrated emission line luminosities, an approximation of the total radiative losses
gives good results, provided the numerical resolution suffice to minimize numerical dissipation effects
\citep{Raga07,TA08}.

Even though these results were obtained in the 1D limit, nonetheless they can serve as a guideline for 
multidimensional case, where additional physical effects, such as rotation \citep{BA02},
can affect the shock evolution. 
\citet{TE09b} and \citet{TE09c} have carried out
preliminary studies of the evolution of 2D shocks traveling along a jet deriving synthetic emission maps,
full synthetic spectra and position-velocity (PV) diagrams of single lines.

Previous theoretical investigations (e.g., \citet{SMK79}) focused on stationary shock models where 
strong shocks were able to produce, via the UV radiation emission, the ionization of the pre-shock 
material, affecting the emission properties. Another approach was the one of \citep{HRH87}, that provided
for the interpretation of observational data a set of plane-parallel shock models, including some 
with totally ionized pre-shock medium, with the relative emission line fluxes. It was noticed, at 
that time, that large differences in the emission properties are related to the pre-ionization state
of the pre-shock medium.

In this paper, we consider the
axisymmetric evolution of a train of shocks as they travel along a jet, differently 
from \citet{MA05a} and \citet{TE09a} that studied a single shock. 

These shocks are produced by imposing a sinusoidal perturbation on the jet 
structure, otherwise in radial equilibrium with the external environment.

The use of AMR allows a careful treatment of the post-shock region, providing (at the highest level of
refinement) a minimum grid size corresponding to about 0.02 AU. The emissivity distribution
obtained by numerical simulations with the PLUTO code \citep{MA07,PLUTO-Chombo} is convolved
with a point-spread-function (PSF) similar to the one of the observing instruments for
comparison with observations. As we shall see, a substantial improvement in
reproducing  the observed emission features can
be achieved by introducing a pre-ionization of the jet material. 
Indeed, as recently pointed out \citep{GU11a}, regions surrounding proto-stars are
 subject to the action of X-rays
able to ionize jet material to an important degree that, due to the low 
recombination rates, lasts up to large distances from the jet origin. 

The plan of the paper is the following: In Section 2 we  discuss the initial
equilibrium, perturbation, pre-ionization and parameters and the 
adopted techniques to model the problem; in Section 3 we present the results for
different choice of the parameters; the conclusions are drawn in Section 4.

\section{The model}
%
%
%

Our model consists of a stationary jet model with a superimposed 
time-dependent injection velocity that produces a chain of perturbations
eventually steepening into shock waves. 
In what follows, the fluid density, velocity, magnetic field and thermal 
pressure will be denoted, respectively, with $\rho$, 
$\vec{v}=(v_r,v_\phi,v_z)$, $\vec{B}=(B_r, B_\phi, B_z)$ and 
$p$. 
The gas pressure depends on the plasma density $\rho$, temperature $T$ and 
composition through the relation $p=\rho k_BT/(\mu m_H)$, where $\mu$ is the
mean molecular weight and $k_B$ is the Boltzmann constant.

Simulations are carried out by solving the time-dependent MHD equations 
in cylindrical axisymmetric coordinates $r,z$:
\small
\begin{equation}\begin{array}{rcl}
\DS \pd{\rho}{t} + \nabla\cdot(\rho\vec{v}) &=& 0 \,,
\\ \noalign{\medskip}
\DS \pd{(\rho v_r)}{t} + \nabla\cdot\left(\rho v_r\vec{v} - B_r\vec{B}\right)
                       + \pd{p_t}{r} &=& \DS\frac{\rho v_\phi^2-B_\phi^2}{r}\,,
\\ \noalign{\medskip}
\DS \pd{(r\rho v_\phi)}{t} + \nabla\cdot\left(r\rho v_\phi\vec{v} - rB_\phi\vec{B} 
                              \right) &=& 0\,,
\\ \noalign{\medskip}
\DS \pd{(\rho v_z)}{t} + \nabla\cdot\left(\rho v_z\vec{v} - B_z\vec{B}\right)
                       + \pd{p_t}{z}    &=& 0\,,
\\ \noalign{\medskip}
\DS \pd{B_r}{t} - \pd{{\cal E}_\phi}{z} &=& 0\,,
\\ \noalign{\medskip}
\DS \pd{B_\phi}{t} + \pd{{\cal E}_r}{z} - \pd{{\cal E}_z}{r} &=& 0\,,
\\ \noalign{\medskip}
\DS \pd{B_z}{t} + \frac{1}{r}\pd{(r{\cal E}_\phi)}{r} &=& 0\,,
\\ \noalign{\medskip}
\DS \pd{E}{t} + \nabla\cdot\left[(E+p_t)\vec{v}-\vec{B}(\vec{v}\cdot\vec{B})
  \right] &=& S_E\,,
\end{array}
\end{equation}
\normalsize
where $\rho$ is the mass density, $\vec{v}=(v_r,v_\phi,v_z)$ is the velocity,
$\vec{B}=(B_r, B_\phi, B_z)$ the magnetic field, 
$p_t = p+\vec{B}^2/2$ denotes the total pressure, 
$\vec{\cal E}=-\vec{v}\times\vec{B}$ is the electric field
and $E$ the total energy density:
\begin{equation}
E = \frac{p}{\Gamma - 1} + \rho\frac{\vec{v}^2}{2} + \frac{\vec{B}^2}{2}\,,
\end{equation}
with $\Gamma = 5/3$ the specific heat ratio.
Also, $F_{ij} = \rho v_iv_j - B_iB_j$ are the flux dyad components
and $\vec{q}=(E+p_t)\vec{v}-\vec{B}(\vec{v}\cdot\vec{B})$ is the 
energy density flux.
The source term $S_E$ accounts for radiative losses and is 
directly coupled to the ionization network described in
\cite{TA08},  
\begin{equation}\label{eq:rhoX}
  \pd{(\rho X_{\kappa,i})}{t} 
          + \frac{1}{r}\pd{(r\rho X_{\kappa,i}v_r)}{r} 
          +            \pd{(\rho X_{\kappa,i}v_z)}{z} = \rho S_{\kappa,i}
\end{equation}
where $\kappa$ and $i$ identify the element and its ionization stage,
respectively, and $S_{\kappa,i}$ is a source term accounting for ionization
and recombination processes.
Given the range of temperature and density, we include the first three 
ionization stages of C, O, N, Ne, S besides hydrogen and helium.

Numerical simulations have been performed in the computational domain 
defined by $r\in [0,400]$ and $z\in[0,1200]$ AU covered by a base grid of 
$128\times 384$ cells, with 6 additional levels of refinement with consecutive 
grid jump ratios of $2:2:4:2:2:2$, thus yielding an effective resolution of 
$16384\times 49152$ cells.
Computations are performed using the AMR version of the PLUTO code with the 
HLLC Riemann solver together the spatially and temporally second-order 
accurate MUSCL-Hancock scheme. 
See \cite{PLUTO-Chombo} for a detailed description of the code and
implementation methods.

\subsection{Model Parameters and Simulation Cases}
%

\begin{deluxetable}{cccc}
\tablecaption{Definition of the simulation sets and corresponding parameters.}
\tablewidth{0.75\columnwidth}
\centering
\tablehead{\colhead{Set} & \colhead{$n_H$} & \colhead{$v^{\max}_{\phi}$(Km/s) } & 
\colhead{$B^{\max}_{\phi}$ ($\mu$G)} }
\startdata
$A (L_X,\tau,\delta v)$ & $10^4$       &   0                      &   152.3   \\ \noalign{\medskip}
$Ah(L_X,\tau,\delta v)$ & $10^4$       &   0                      &     -     \\ \noalign{\medskip}
$B (L_X,\tau,\delta v)$ & $10^4$       &  10                      &   422.4   \\ \noalign{\medskip}
$C (L_X,\tau,\delta v)$ & $5\cdot10^4$ &   0                      &   152.3   \\ \noalign{\medskip}
$D (L_X,\tau,\delta v)$ & $5\cdot10^4$ &  10                      &   422.4   \\ \noalign{\medskip}
$E (L_X,\tau,\delta v)$ & $5\cdot10^4$ &  15                      &   610.3  
\enddata
\tablecomments{
Different simulation cases are distinguished by the hydrogen density $n_H$, 
peak rotation velocity $v^{\max}_\phi$ and magnetic field $|B^{\max}_{\phi}|$ 
respectively given in the second, third and fourth column.
Each set defines a family of models with varying X luminosity $L_X$ of the 
central object, period and amplitude of the perturbation $\tau$ and $\delta v$.
In all simulation cases, the jet radius, temperature, velocity and density 
contrast are the same and equal to $r_j=20\,{\rm AU}$, $T_j=2 \, 500 \,K$, 
$v_j=110\,{\rm km/s}$ and $\eta = 5$, respectively.
}
\label{tab:param}
\end{deluxetable}

A cylindrical jet equilibrium model is constructed by first prescribing 
radial profiles for density, velocity, magnetic field and then 
by solving the radial balance momentum equation for the gas pressure. 
The details of this equilibrium configuration are
outlined in the Appendix \ref{sec:equil}.
The resulting radial profiles define a family of jet models 
characterized by the hydrogen number density $n_H$, 
longitudinal velocity $v_j$, temperature $T_j$, 
jet to ambient density contrast $\eta=n_H/n_a$ and peak rotation
velocity $v^{\max}_{\phi}$.
In the present context we restrict our attention to purely toroidal
configurations and leave models with helical magnetic fields 
(i.e. $B_z\ne0$) to forthcoming studies.
Since the ambient temperature is prescribed to be $T_a=1 \, 000$ K, the maximum 
value of $B_\phi$ is not a free parameter but depends on the rotation velocity.

Finally, the parameter that controls the degree of pre-ionization of 
the jet material at the base of the jet is the X-ray luminosity $L_X$ 
of the central object, for which the ionization at photoionization 
equilibrium is computed as explained in \S\ref{sec:preion}.

Along with the equilibrium magnetized models we also consider
purely hydro configurations that, due to an over-pressurized beam, 
cannot establish equilibrium with the environment. 
In this case a conical structure is formed during the propagation.

\begin{figure}
\centering
\includegraphics[width=0.45\textwidth]{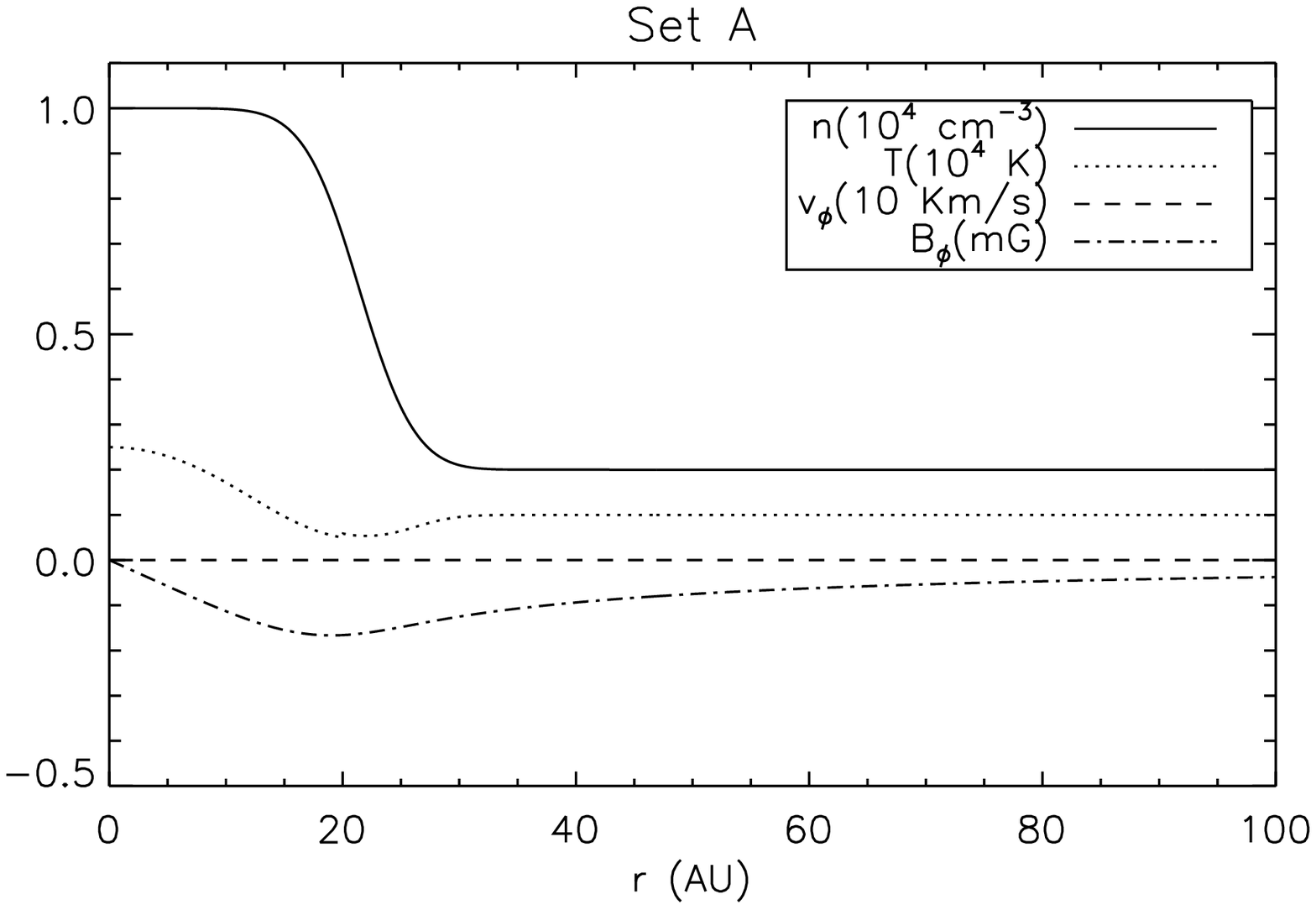}
\includegraphics[width=0.45\textwidth]{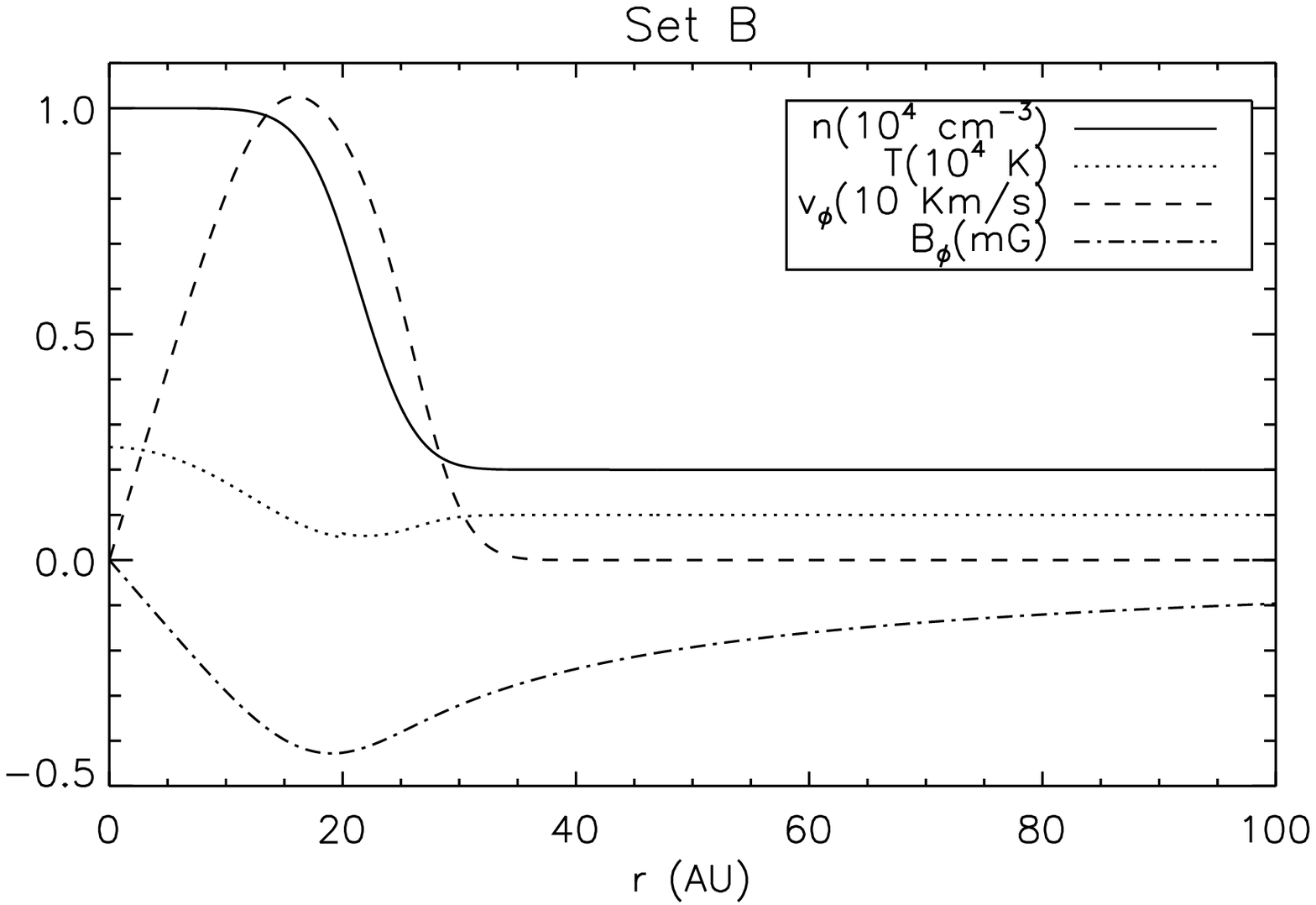}
\caption{\footnotesize Radial equilibrium profiles for set A (top panel) and
         set B (bottom panel). In each panel we plot density 
         (in $10^4\,{\rm cm}^{-3}$, solid line), temperature (in $10^3$ K, 
         dotted line), azimuthal velocity (in $10$ km/s, dashed line) and 
         magnetic field (in $10^{-3}$ Gauss, dash-dotted line).}
\label{fig:equil_profile}
\end{figure}

In the simulations reported here we set the initial jet temperature, 
velocity and density contrast to the values $T_j=2500\,{\rm K}$,
$v_j=110$ Km/s and $\eta = 5$, respectively. 
Table \ref{tab:param} summarizes the chosen set of simulation cases
while we plot in Fig (\ref{fig:equil_profile}) the radial profiles
for density, temperature, velocity and magnetic field.
Within each set (labeled by a capital letter), the X luminosity of the 
central object, the period and amplitude of the perturbation are 
allowed to vary.

Set $A$ is characterized by no rotations and a relatively weak magnetic 
field and density.
As a special case, we also include set $Ah$ consisting of purely-HD 2D simulations. 
In these cases, the conical expansion favors the formation of a decreasing density 
along the longitudinal direction.
Set $B$ has stronger rotation and (consequently) magnetic field. 
Sets $C$ and $D$ are identical to $A$ and $B$ (respectively) except that the beam
is five time heavier. Finally, the last set $E$ has a maximum rotation 
velocity $v^{\max}_{\phi}=15$ Km/s and peak magnetic field of $610\, \mu$G.

\subsection{Initial perturbation}
%

In previous works on astrophysical jets, we have employed a special definition of the initial 
perturbation (described in \citet{MA05a}), imposing conditions that led to the
formation of only one shock propagating along the jet beam, instead of the usual pair of 
forward-reverse shocks. This approach was preferred because it allowed a higher level of control 
on the energy dissipation areas and an easier parallel between the perturbation parameters and
the characteristics of the forming shockwave.

In the present work however, a time-dependent velocity 
fluctuation is prescribed at the boundary (after a steady configuration has been reached) as:
\begin{equation}
\label{eq:sinpert}
  \delta v_z= A\sin\left(\frac{2\pi}{\tau} t \right) 
\end{equation}
where $\tau$ is the period of perturbation (in years).
This choice is justified by two main reasons: \\
(1) the formation of the pair of forward-reverse shocks elongates the high intensity line emission 
area and leads to a better agreement with the morphology of the observed emission knots, and \\ 
(2) our aim of approaching simulation results to observational data benefits from less strict 
conditions on the perturbation parameters.

Moreover, we limit ourselves to three perturbation periods since the conditions in which 
the second and the third shock propagate are quite similar.

\subsection{Pre-ionization fraction}
\label{sec:preion}
%

We analyze the effect of the jet base irradiation by X-rays coming from the central TTauri star.
Our goal is not to model this region in detail, but is limited to gain information on reasonable values
of the ionization of the jet medium at the distance where observations and our
simulations start, i.e. at $r_{\rm s}=0^{\prime\prime}.1$ corresponding to $\sim 2 \times 10^{14}$ cm.
Detailed numerical calculations of the combined dynamical, heating-cooling and photo-ionization
processes in YSO jets are under way and will be published in a forthcoming paper.

Proto-stellar objects show X-ray luminosities $10^{28}-10^{32}$ ergs s$^{-1}$, 
depending on their mass and possibly originating from the magnetized stellar corona 
\citep{GA00,PN05}, with possible contributions from the jet itself, as discussed recently by
\citet{SK11} for RY Tau - HH 938 and by \citet{GU11b} for DG Tau. 
The interaction of a X-ray photon, in the keV energy range,
with an atom or molecule results in the production of a fast photoelectron,
the primary, that in turn generates, collisionally, a deal of secondary electrons
\citep{GA97}. We follow the treatment by \citet{SA02}, that ignores
the contribution of the primary electrons and considers the dominant secondary
electrons only. We write the energy input ${\cal H_{\rm X}}$ by X-rays (energy  per unit volume 
per unit time) and the photo-ionization rate $\zeta_{\rm X}$ as:
\begin{equation}
{\cal H_{\rm X}}= 
\frac{n_{_{\rm H}}(r)}{4 \pi r^2}\int_{E_0}^\infty L_{\rm X}(E)
\sigma_{\rm pe}(E) \ e^{-\tau_{\rm X}} \ y_{\rm heat} \ dE \;,
\label{eq:heat}
\end{equation}

\begin{equation}
\zeta_{\rm X}= \frac{1}{4 \pi r^2}\int_{E_0}^\infty 
\frac{L_{\rm X}(E)}{\epsilon_{\rm ion} }
\sigma_{\rm pe}(E) \ e^{-\tau_{\rm X}} \ \ dE \;.
\label{eq:yon}
\end{equation}
In the expression above $L_{\rm X}(E)$ is the energy dependent X-ray luminosity, 
$E_0(=0.1$ keV) is the low-energy cutoff,  $\sigma_{\rm pe}(E) $ is the cosmic photoelectric
absorption cross section per H nucleus,
$y_{\rm heat} $ is the absorbed fraction of the X-ray flux, 
$\epsilon_{\rm ion} $ the energy to  make an ion pair, and $r$ is the optical path
in spherical symmetry.
Since $y_{\rm heat} $ and $\epsilon_{\rm ion} $ (given by \citealt{SA02}) can be
considered nearly independent of energy, we have
\begin{equation}
{\cal H_{\rm X}}= n_{_{\rm H}}(r) \ y_{\rm heat} \ 
\epsilon_{\rm ion} \ \zeta_{\rm X} \;,
\label{eq:heat1}
\end{equation}
 where (\citealt{SS85})
\begin{equation}
 \frac{1}{\epsilon_{\rm ion}}= \frac{y_{\rm H}}{I(H)} + 
\frac{y_{\rm He}}{I(He)}  \;,
\label{eq:heat2}
\end{equation}
with
$$
y_{\rm H}=0.3908 \ (1-x_{\rm e}^{0.4092} )^{1.7592} \;,
$$
$$
y_{\rm He}=0.0554 \ (1-x_{\rm e}^{0.4614} )^{1.666} \;.
$$
In the above relationships
$I(H)$ and $I(He)$ are the ionization potentials of $H$ and $He$,
$x_{\rm e}$ is the hydrogen  ionization fraction, and
$$
y_{\rm heat} = 0.9971 \ [1-(1-x_{\rm e}^{0.2663})^{1.3163}] \;
$$
specifies the heating fraction. 

The X-ray optical depth  $\tau_{\rm X}$ can be written:
\begin{equation}
\tau_{\rm X} = \sigma_{\rm pe}(k T_{\rm X})N \, , \; N= \int_{0}^r n_{\rm H} dr^\prime \; ,
\label{eq:tau}
\end{equation}
where $\sigma_{\rm pe}(E)= \sigma_{\rm pe}(k T_{\rm X}) ({keV} / E)^p$ and 
$\sigma_{\rm pe}(1 {\rm keV}) =2.27 \times 10^{-22} \ {\rm cm}^2$, $k T_{\rm X}=1 {\rm keV}$
and the exponent $p=2.485$ is for solar abundances.

Note that for a thermal spectrum the ionization rate (Eq. \ref{eq:yon}), becomes
\begin{equation}
\zeta_X= \frac{L_{\rm X} \sigma_{\rm pe}(k T_{\rm X})}{4 \pi r^2 
\epsilon_{\rm ion}} 
\int_{\xi_0}^\infty \xi^{-p}\exp{[-(\xi+\tau_{\rm X}\xi^{-p}}]
 \ d\xi \;.
\label{eq:yon1}
\end{equation}
where $L_{\rm X}$ is the total X-ray luminosity  and $\xi=E/kT_{\rm X}$.

We consider the region close to the inner disk, where the disk-wind jet component is originated,
and above the extended stellar atmosphere, where the stellar-wind jet component
is being launched (see discussion in \citet{MA09b}). The medium there is heated and ionized
 by a X-ray flux of luminosity $L_{\rm X}$. This region extends from
a distance $r=R_{\rm X}\sim 10^{12}$ cm ($\sim 
10 \ {\rm R}_\odot$) from the star, i.e. the stellar corona outer radius,
up to about 1 AU, i.e. the inner disk. The radial velocities there are small enough and the ionization, 
recombination, heating and cooling timescales fast enough that we can assume energetic and 
ionization/recombination equilibria:
\begin{eqnarray}
{\cal H_{\rm X}}-{\cal L} =0 \;,
\label{eq:ener} \\
(c_{\rm i}+ \frac{\zeta_X}{n_{\rm e} }) f_{\rm n} - c_{\rm r} (1- f_{\rm n}) =0 \;,
\label{eq:ion}
\end{eqnarray}
with $f_{\rm n}$ the number fraction of  neutral hydrogen atoms, 
$n_{\rm e}=n_{\rm H}(1-f_{\rm n}+Z)$ the electron density,
$n_{\rm H}$ the total hydrogen density and $Z$ (=0.001) the
metal abundance by number, $c_{\rm i}$, $c_{\rm r}$ are
the ionization and recombination rate coefficients, respectively (see \citet{DS03}),
and ${\cal L}$ represents the energy loss term (energy  per unit volume per unit time).
The loss term is modeled according to the SNEq cooling model by \citet{TE09a}.

If we assume a very moderate X-ray luminosity $L_{\rm X}=10^{29}$ ergs s$^{-1}$
and a particle density of $ 10^6$ cm$^{-3}$, at $r=R_{\rm X}$ we obtain, according to Eqs. \ref{eq:ener}
and \ref{eq:ion}, equilibrium temperature $\approx 12,000$ K and ionization fraction $\approx 40$\%
close to the jet axis and drops to about 5\% at 1 AU, at the jet initial lateral border. The 
ionization/recombination timescales are of the order of months, while the heating/cooling ones
are about an order of magnitude smaller. The matter is then funnelled into the jet by dynamical 
and MHD processes, expands and accelerates reaching velocities of $100-200$ km s$^{-1}$ in a 
few AUs \citep{ZA07,TA09}. One may expect a substantial drop in temperature by cooling, but the
ionization fraction, due to long recombination timescale, $t\sim 1/(c­_r n_e)$, would remain 
 close to the equilibrium one. Thus, the assumption of a residual ionization fraction in the 
central spine of the jet of about 10-20\% at $0^{\prime \prime}.1$ is a quite reasonable one.

\subsection{Post-processing and data analysis}
%

The output from numerical simulations,
that include the chemical/ionization network and radiative cooling losses, cannot be directly
compared with observations. Density, velocity and ionization fraction distribution
must be in fact transformed into surface brightness maps, line ratios and Position-Velocity diagrams
in a post-processing phase.

The first step in this process is the computation of 2D emissivity maps at wavelengths corresponding
to atomic transitions of interest, selected by the user. In the 5-level atom model considered by the 
cooling treatment implemented in the PLUTO code, there are a few hundred selectable emission lines.
For these computations, the ionization state of the matter and the temperature in each simulation cell
must be known. The simulation code PLUTO delivers the detailed ionization state for the 
atomic species H, He, C, N, O, Ne and S. The temperature is computed from the pressure,
density and ionization state in each cell.

The second step is the 3D emissivity integration, in cylindrical
symmetry, done by rotating the 2D emissivity maps previously obtained around the $z$ axis.
The 3D structure is then projected onto a plane perpendicular on the line of sight (the emitted power
in each emission line is integrated over lines parallel to the line of sight), in order to 
obtain a surface brightness map similar to the ones observed. A simulation of the effects of the
PSF of the instrument is also added, usually the simulations having much higher
resolutions than the observational data (in order to capture the physics within). The 
PSF assumes a Gaussian form, with user-defined half-width $\sigma$. For the 2D surface brightness maps
presented in this work, a PSF that is roughly 1/4 of the one of HST was employed (HST has a resolution
of approximately $0^{\prime\prime}.1$, that means 14 AU at the distance of Taurus-Aurigae where the sources are located). 
We have chosen to use this smaller PSF in order to have, at this stage, a better resolution of the 
output jet structures. In drawing the plots of line ratios and surface brightness along the axis of the simulated jet, 
the resolution was reduced to approximately that of HST.

\begin{figure}[!h]
\centering
\includegraphics[width=0.5\textwidth]{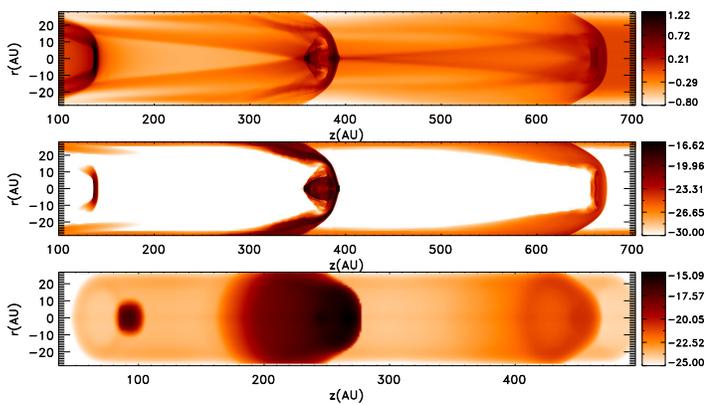}
\caption{\footnotesize Top-panel: Logarithmic density map from PLUTO output; 
                       Middle: Emissivity in [SII]6716\AA in units of erg cm$^{-3}$s$^{-1}$, logarithmic map;
		       Bottom: Surface brightness map in erg cm$^{-2}$arcsec$^{-2}$s$^{-1}$, logarithmic, angle jet - LoS 45 deg. }
\label{fig:postproc}
\end{figure}

The two steps leading from the PLUTO output data to simulated maps of surface brightness 
are illustrated in Fig. \ref{fig:postproc}. 

After the second step of post-processing, a longitudinal or transversal slit of arbitrary 
size can be defined on the computed surface brightness map,
 used to compute synthetic spectra and position-velocity 
diagrams. The synthetic spectra include the natural and Doppler line broadening, and consist 
of all emission lines selected for processing, with customizable spectral range and resolution. 
The resulting position-velocity (PV) diagrams can be directly compared to the ones derived from 
observations. PV diagrams taken with a slit parallel to the jet axis and stepped across the
jet or a slit perpendicular on the jet axis are particularly useful for simulations that include 
the rotation of the jet. This is expected from models of jet generation, and indications of 
rotation have been detected in several microjets in recent works \citep{BA02,WA05,CA04,CA07}.

It is also possible to extract velocity channel maps in custom velocity channels and emission 
lines, to be compared with observations. These velocity channel maps are of paramount importance
in the investigation of jet structure.

\section{Results}
%
%
%

We discuss the results of the numerical simulations and compare these with observations of
emission knots of the three sources, 
for which high-quality observational data are available in the literature.
The jets obtained with the numerical simulations have been projected at an
angle of 45 degrees with the line of sight, 
taking as a reference the case of the RW Aurigae jets.

\subsection{Shocked jet emission}
%

In the simulation set $A$, the equilibrium of a cylindrical jet is guaranteed by the
toroidal component of the magnetic field vector and the density along
the jet remains uniform, thus the first shock propagates in a constant density environment.
On the contrary, the second and the third shocks in the array
 travel in the decreasing density zone following the propagation of the previous shock, ensuring
a longer time-span for intense line emission.
Indeed, following the evolution of the shocks over time, one can notice the different behaviour 
of the second shock with respect to the first one, being brighter over a larger distance. 

\begin{figure} [!h]
\begin{center}
\begin{tabular}{c}
 \resizebox{80mm}{!}{\includegraphics{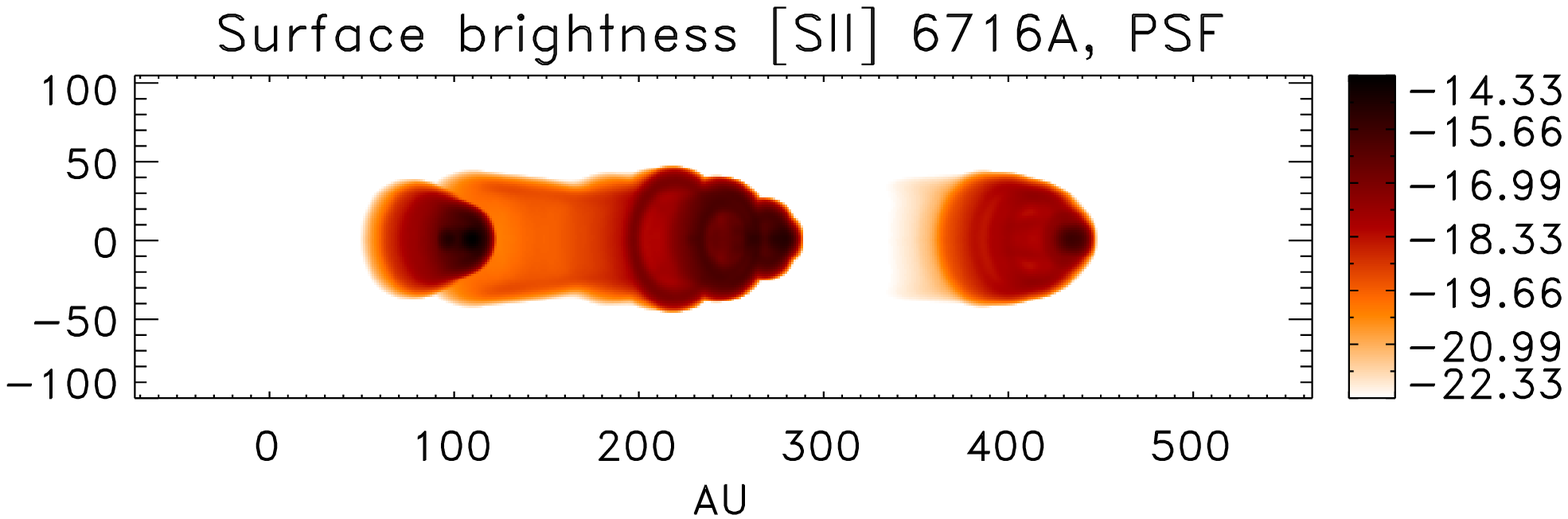}} \\
 \resizebox{80mm}{!}{\includegraphics{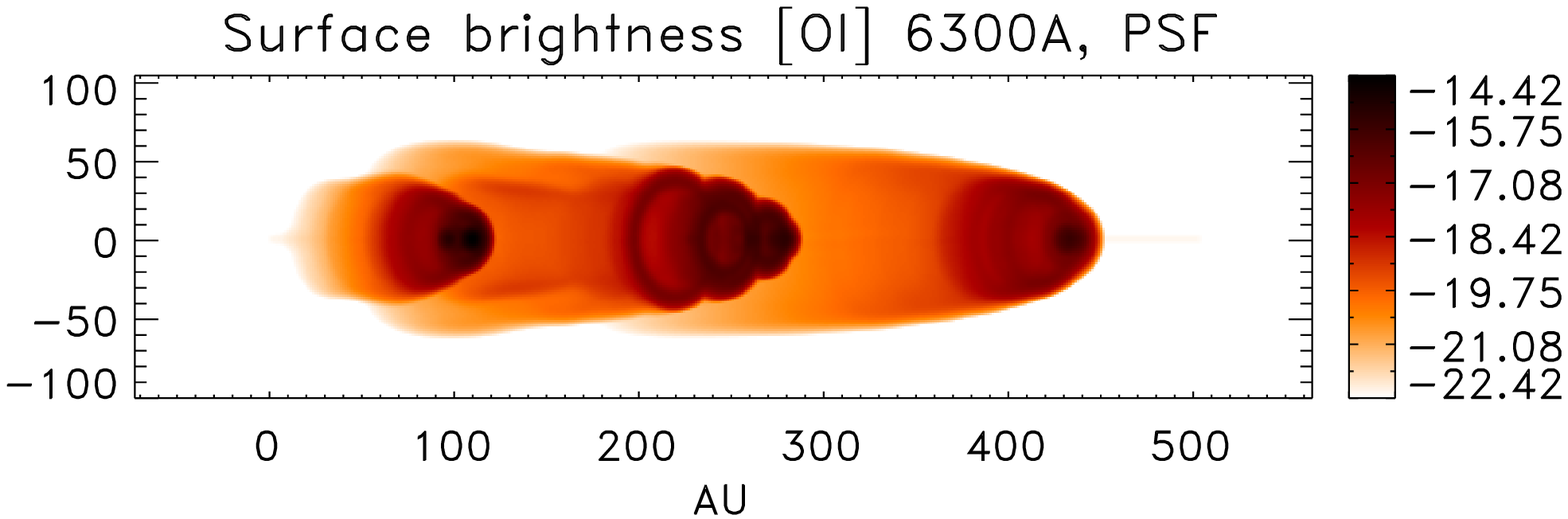}}       \\
 \resizebox{80mm}{!}{\includegraphics{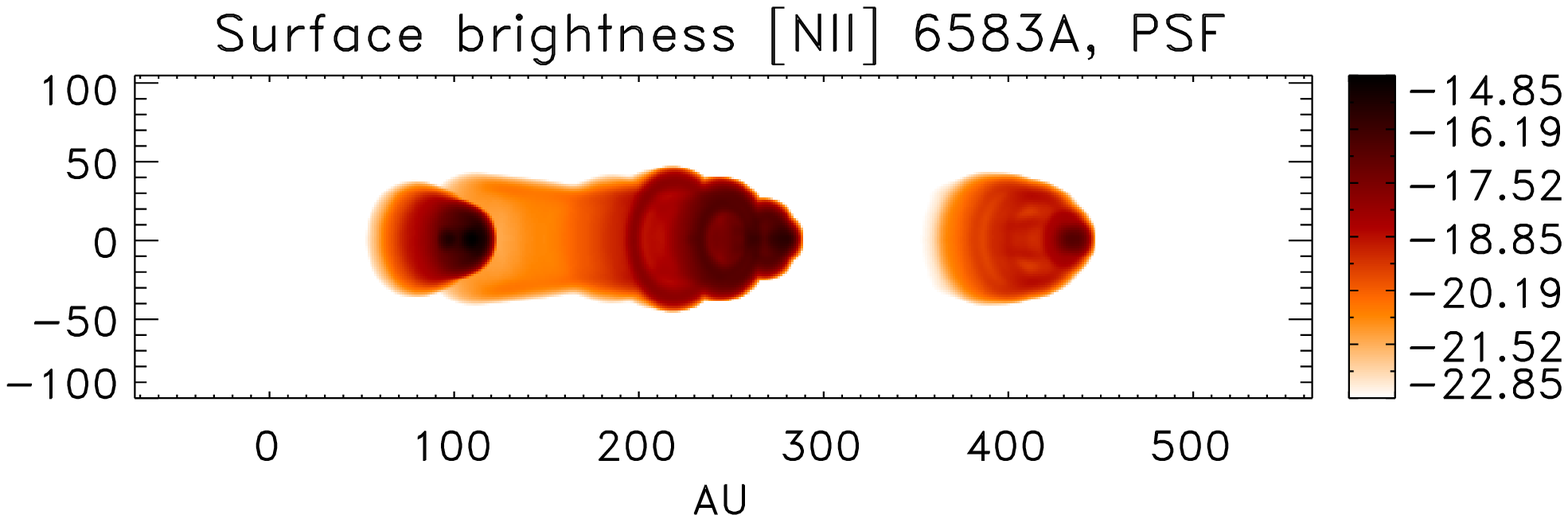}}
\end{tabular}
\caption{Simulation in $A$ configuration, no pre-ionization, perturbation amplitude 50km/s, period 10 years, 
         surface brightness in [SII]6716\AA (top), [OI]6300\AA (middle) and [NII]6583\AA, 
         units of erg cm$^{-2}$arcsec$^{-2}$s$^{-1}$, log10 maps.}
\label{fig:bright_noion}
\end{center}
\end{figure}

Fig.\,\ref{fig:bright_noion} shows surface brightness 
maps in three emission lines from [SII], [OI], and [NII] respectively, for a simulation type 
$A$. In this case the jet variability period is 10 years, the perturbation amplitude 50km/s and the 
temperature of the jet material 2\,500K (Hydrogen mostly neutral before the shock). We can see 
in this figure the  sharp decrease of the brightness after the peak of about four orders of 
magnitude over a distance of 50AU. This leaves large dark spaces between emission knots, that are not seen in 
observations. We note that an attempt to alleviate this problem by diminishing the time periodicity 
of the perturbations that evolve in shock waves, lead to a decrease in the maximum 
knot brightness, explained by the lower mass flux entering each shock.

\begin{figure} [!h]
\begin{center}
\begin{tabular}{c}
 \resizebox{80mm}{!}{\includegraphics{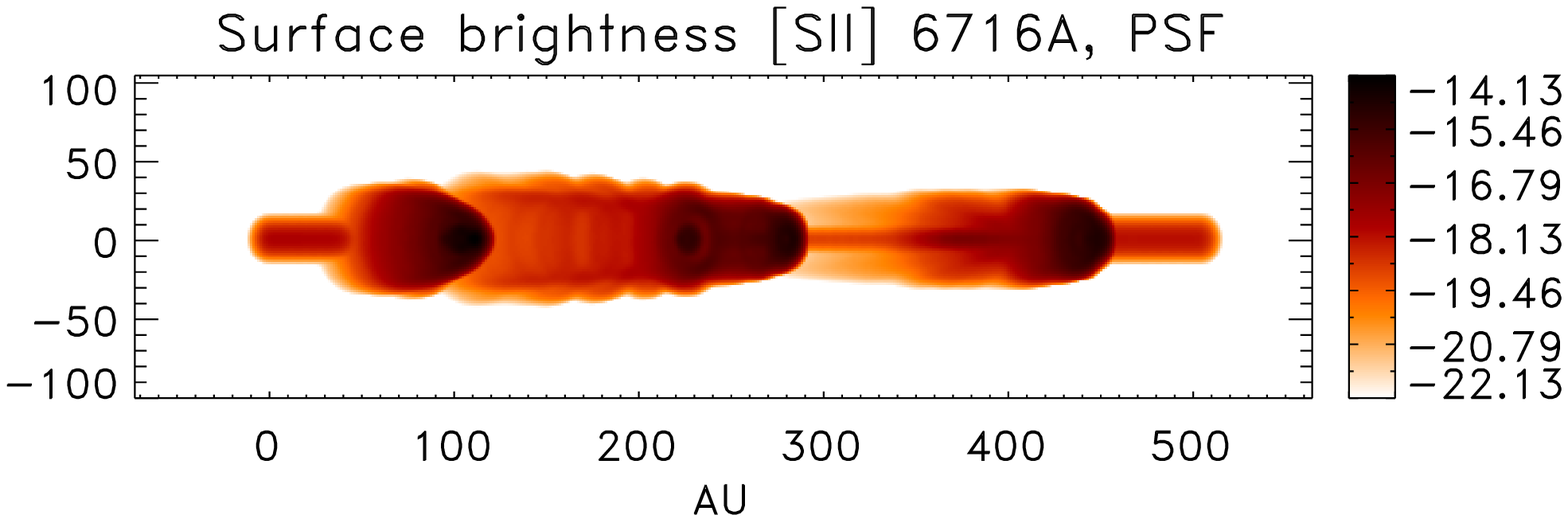}} \\
 \resizebox{80mm}{!}{\includegraphics{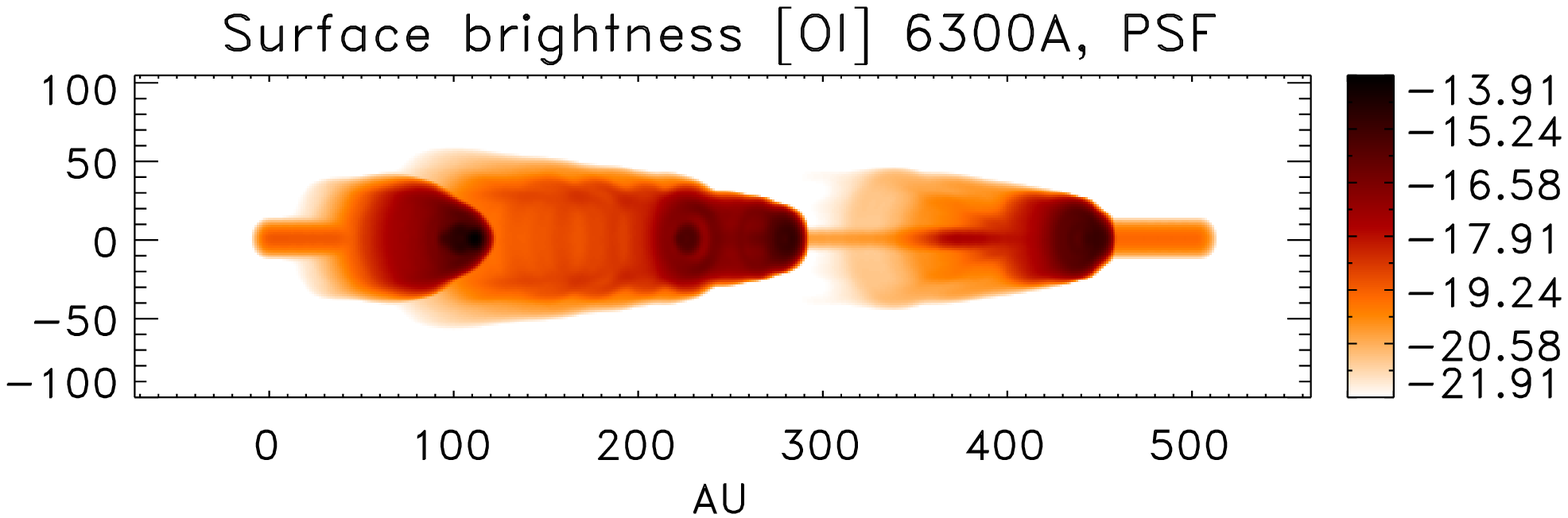}} \\
 \resizebox{80mm}{!}{\includegraphics{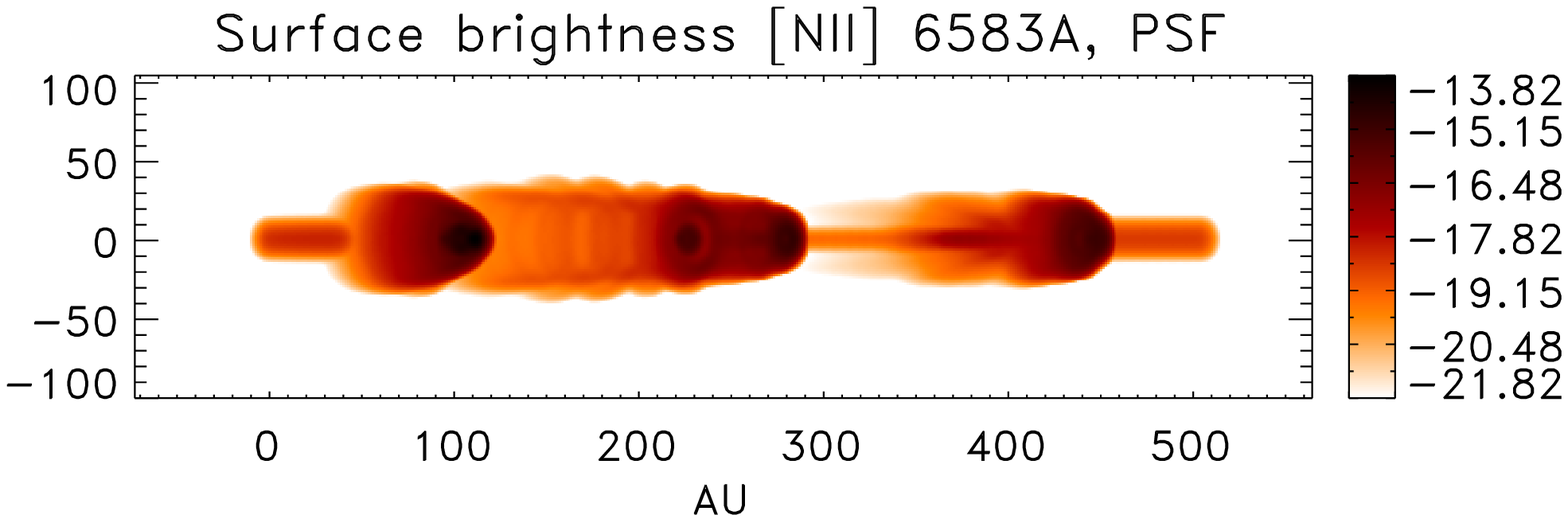}}
\end{tabular}
\caption{Simulation in $A$ configuration, pre-ionization 19\%, perturbation amplitude 50km/s, period 10 years, 
         surface brightness in [SII]6716\AA (top), [OI]6300\AA (middle) and [NII]6583\AA, 
         units of $erg\cdot cm^{-2}arcsec^{-2}s^{-1}$, log$_{10}$ maps.}
\label{fig:bright_preion}
\end{center}
\end{figure}

When an X-ray-induced pre-ionization of the pre-shock medium is considered
(about 19\% in Hydrogen), Fig.\,\ref{fig:bright_preion}, the
emission areas behind the shocks are extended compared to previous case, Fig.\,\ref{fig:bright_noion} 
(in the figures being presented the same moment in the evolution), and the maximum values 
of the brightness are higher as well.
This configuration provides surface brightness maps more similar to observational data, 
with elongated emission knots because of the higher background ratio of ionized elements
in the jet material.

Moreover, the presence of pre-ionization leads to the increase in the peak surface brightness
with factors between 2 and 4. This is due to the fact that a pre-existing increased number
of free electrons fasten the collisional ionization and excitation, enhancing the total
brightness.

\begin{figure} [!h]
\begin{center}
\begin{tabular}{c}
 \resizebox{80mm}{!}{\includegraphics{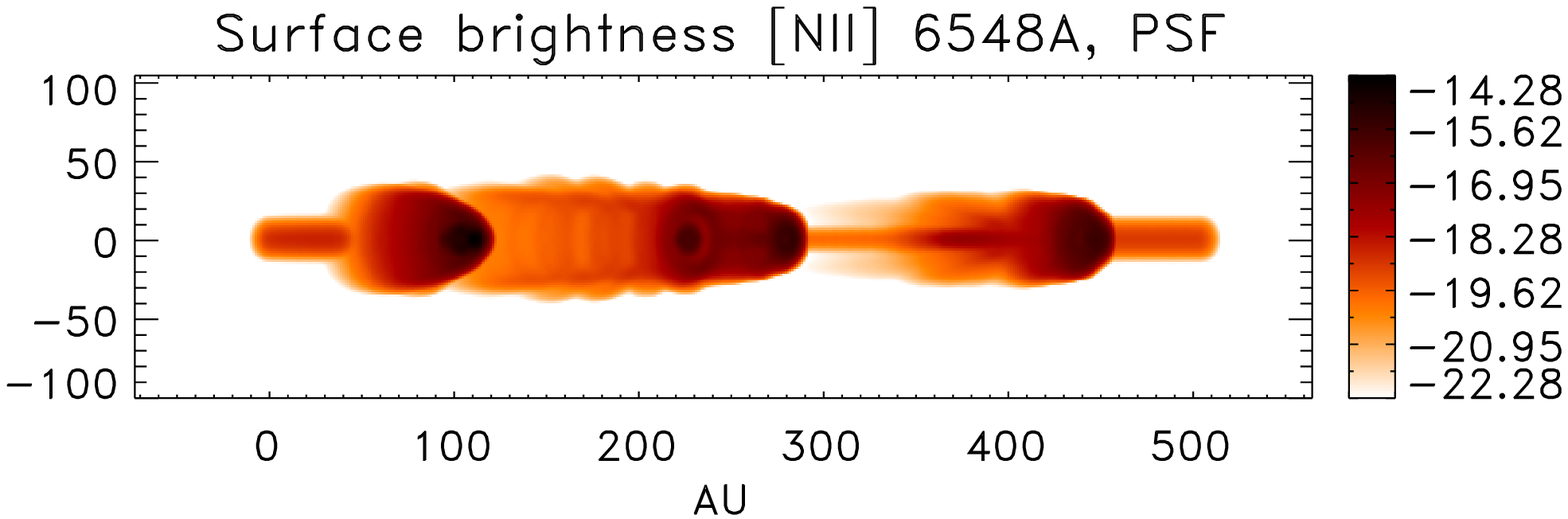}} \\
 \resizebox{80mm}{!}{\includegraphics{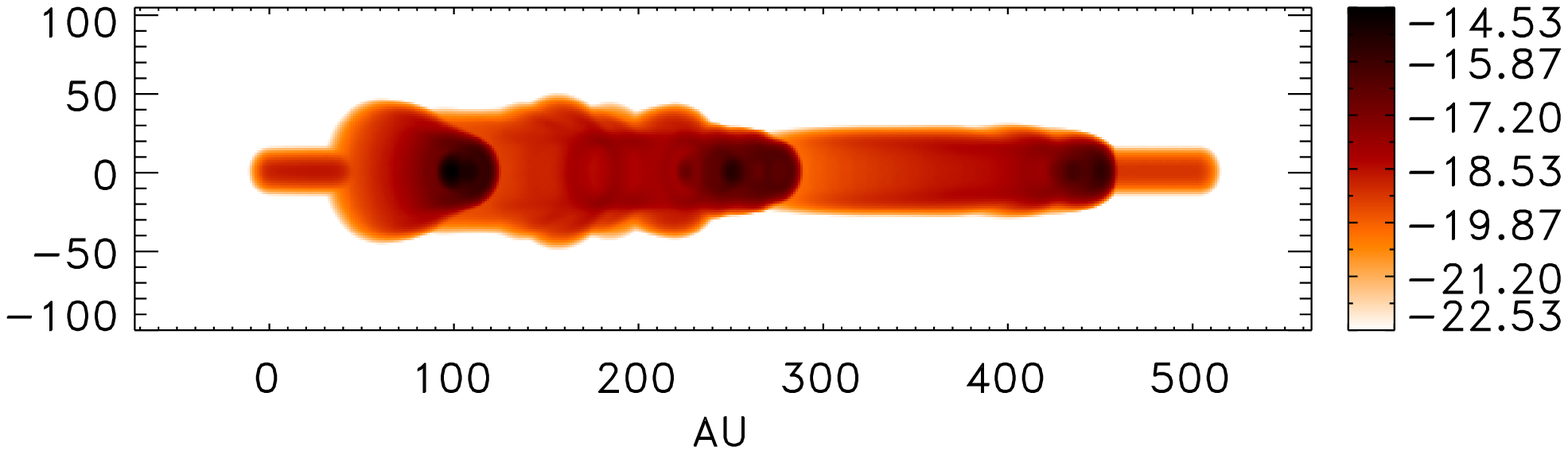}}  \\
 \resizebox{80mm}{!}{\includegraphics{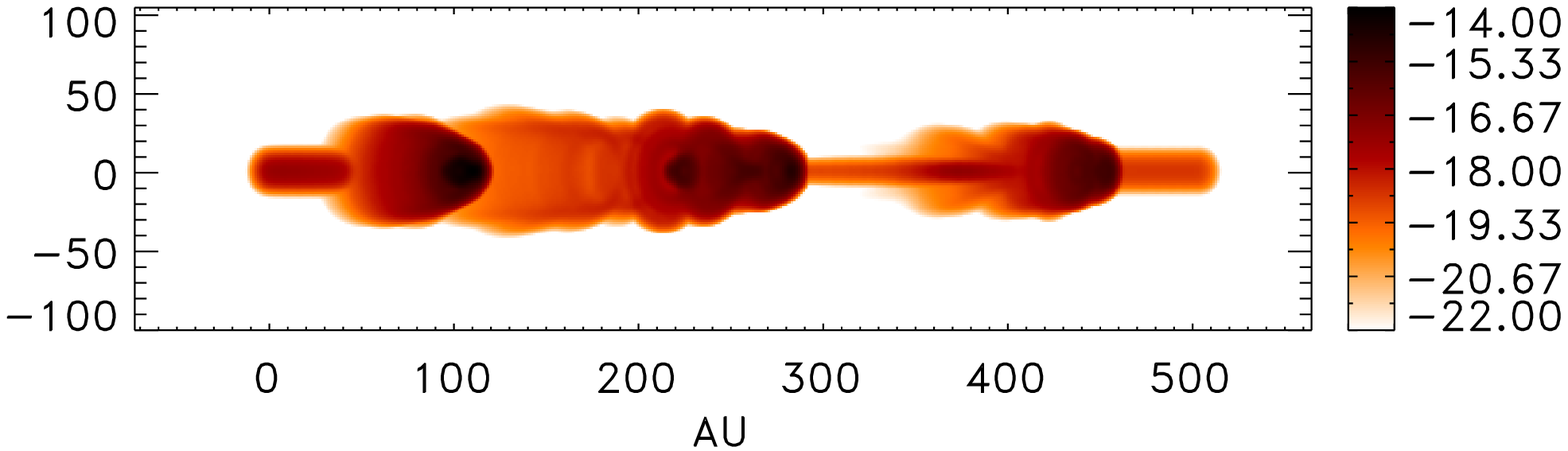}}  \\
 \resizebox{80mm}{!}{\includegraphics{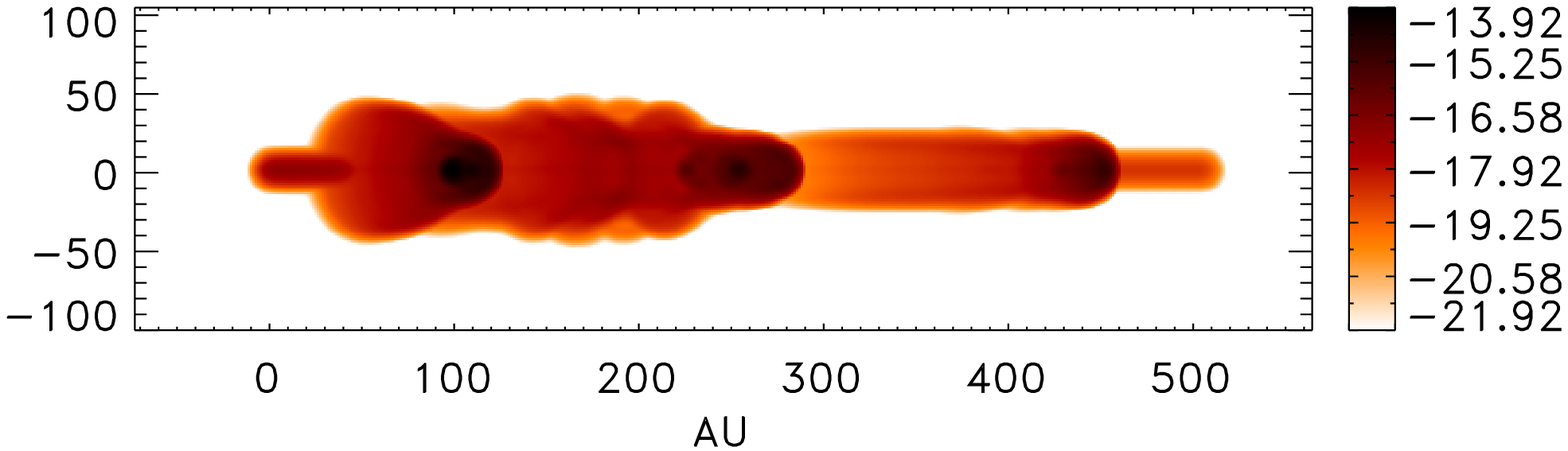}}  
\end{tabular}
\caption{Surface brightness maps in [NII]6548\AA, in four simulation configurations (A, B, C, and D),
         with pre-ionization and the same set of parameters. Units of 
         erg cm$^{-2}$arcsec$^{-2}$s$^{-1}$, log$_{10}$ maps.}
\label{fig:bright_comp}
\end{center}
\end{figure}

In Fig.\,\ref{fig:bright_comp} we show a comparison among the simulated surface brightness of the 
shocked jet in [NII]6548\AA for four different simulation sets (A, B, C, and D from top
to bottom panels), including the pre-ionization of the jet material by X-rays. 
For consistency, the maps are drawn at the same evolutionary stage and the ``variable" 
parameters were set to the same values. 

The top panel shows the surface brightness map for a simulation in setup $A$, 
with rather compact emission knots and low-intensity gaps between them. 
The $B$ simulation (second panel from top) includes the jet rotation with a maximum velocity 
of $10 km\cdot s^{-1}$,
and produces maximum surface brightness lower than in the corresponding $A$ cases, 
but with a reduced decrease in brightness in the regions between two successive emission peaks.
In case $C$ (third panel in Fig.\,\ref{fig:bright_comp}), the propagation of the 
knots is slightly faster with respect to the previous cases, because of the higher density 
($5\cdot 10^4cm^{-3}$ instead of $10^4cm^{-3}$). In addition, the maximum value of the surface 
brightness is higher than in the otherwise very similar results of case $A$.
The results of case $D$ has been obtained setting the
jet rotation at $10 km\cdot s^{-1}$ and density at $5\cdot 10^4cm^{-3}$), 
and from  Fig.\,\ref{fig:bright_comp}, bottom panel, we see that the
morphology of the line emission is similar to the one of case $B$, but with higher 
emission intensities due to the increased amount of mass load of the jet.

The purely hydrodynamic case $Ah$ is characterized by a larger lateral expansion, 
thus both the maximum 
surface brightness and the length of the high-intensity zone result lower than in the
corresponding MHD cases, so it was excluded from the comparison in Fig.\,\ref{fig:bright_comp}.
The results in the $E$ cases were very similar to the ones obtained in the $D$ setup,
thus not displayed.

\subsection{Comparison with observations}
%

\subsubsection{Observational constraints}

We refer to Hubble Space Telescope (STIS instrument) observations of RW Aurigae jets, \citet{MA09};
for DG Tau, \citet{BA02} and \citet{LA00}; and for HH30, \citet{HM07}.

\begin{figure}[!h]
\centering
\includegraphics[width=0.5\textwidth]{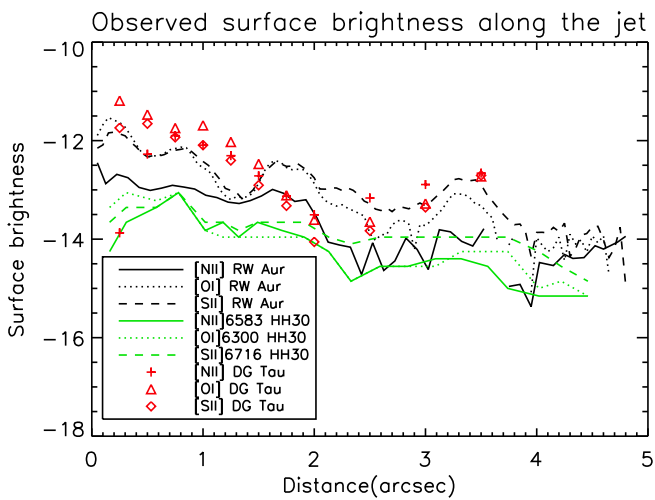}
\caption{\footnotesize Surface brightness in along the jets in units of 
                       erg cm$^{-3}$arcsec$^{-2}$s$^{-1}$, logarithmic plot on the jet axis from 
                       observations of RW Aurigae redshifted jet, DG Tau and HH 30. }
\label{fig:bright_obs}
\end{figure}

In Fig.\,\ref{fig:bright_obs}, we show the observed surface brightness along the jet axis 
in the three emission doublets of [OI] (6300\AA and 6363\AA), [NII] (6548\AA and 6583\AA) and 
[SII] (6716\AA and 6731\AA) for the three sources quoted above. Hereafter, where
no wavelength is specified, the square brackets notation refers to the sum of both lines of the 
respective doublet.

One can note the overall higher brightness of the three emission doublets for DG Tau, 
in agreement with both the higher Doppler velocities measured for this source and the
presence of an X-ray emission discovered by the Chandra Observatory (e.g. \cite{SS08}), 
as possibly indicative stronger shock waves. 
{\emph Working in the approximation of optically-thin
plasma, the higher values for the RW Aurigae redshifted jet with respect to HH30 jet, despite the 
similar flow and shock velocities, may be explained by the higher declination angle of the former
with respect to the line of sight and the different toroidal magnetic field strength.
}

\subsubsection{Surface brightness}

\begin{figure}[!h]
\centering
\includegraphics[width=0.5\textwidth]{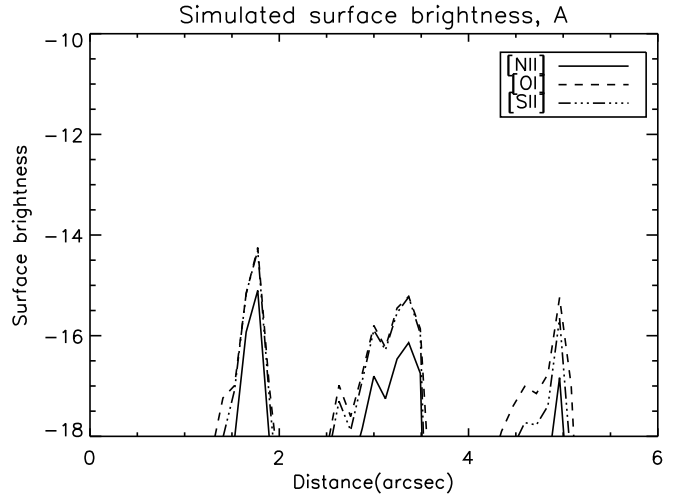}
\caption{\footnotesize Surface brightness, doublets of [SII], [OI] and [NII] in units of 
                       erg cm$^{-3}$arcsec$^{-2}$s$^{-1}$, logarithmic plot on the jet axis from 
                       simulation type $A$, with no pre-ionization. }
\label{fig:brightA}
\end{figure}

As discussed in the previous section, the surface brightness variation with distance along
the jet differs depending on the case considered. Without including pre-ionization (i.e.
with the ionization fraction taken in collisional equilibrium 
at 2\,500K ahead of the shocks), the distribution of surface
brightness along the jet (Fig.\,\ref{fig:brightA}) has variations of many orders of magnitude
and lower peak values with respect to the pre-ionized cases (and much lower than observations).

\begin{figure}[!h]
\centering
\includegraphics[width=0.5\textwidth]{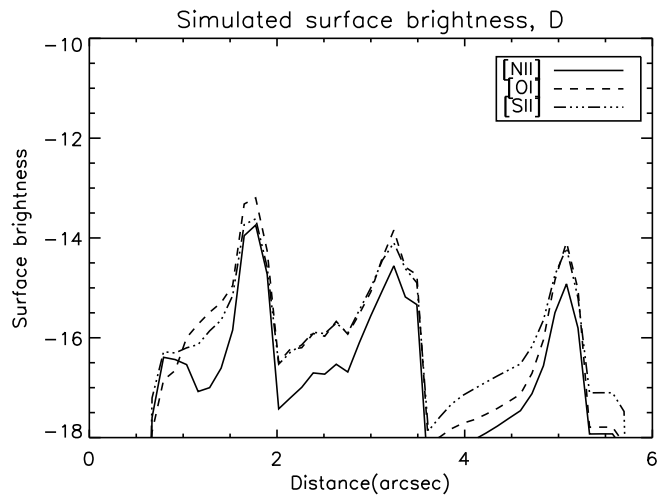}
\caption{\footnotesize Surface brightness, doublets of [SII], [OI] and [NII] in units of 
                       erg cm$^{-3}$arcsec$^{-2}$s$^{-1}$, logarithmic plot on the jet axis from 
                       simulation type $D$, with pre-ionization. }
\label{fig:brightB}
\end{figure}

The rotating jet simulated in configuration $D$ is a good candidate for the comparison with 
observations, the decrease of brightness between the high-intensity being less pronounced than 
in the corresponding non-rotating case ($A$) - see Fig.\,\ref{fig:brightB}.

An important increase in brightness is also important for the comparison with observations -- 
shocks with higher-amplitude perturbations (higher than 50 km s$^{-1}$) are not likely for ``slow" jets such 
as HH30 and RW Aur, so the pre-ionization provides a way of enhancing brightness without 
going with the simulations beyond the most probable parameter range.

The decreasing trend of the peak brightness with the traveled distance from the jet origin
is visible both in simulations and observations: at angular distances larger than
 $2^{\prime\prime}$, the decrease is approximately one order
of magnitude (Figs. \ref{fig:bright_obs} and \ref{fig:brightB}).
This suggests that the knots observed in many jets (e.g. HH 34 and HH 111)
at distances of a few tens of arcseconds from the source
are likely to arise from other mechanisms, i.e. jets shear-layer
instabilities. 

\subsubsection{Line emission ratios}

The line emission ratios are indispensable ingredients
in methods for deriving the physical parameters of space 
plasmas from observations. In the case of stellar jets - the forbidden emission doublets of [SII], [OI]
and [NII] between 6 and 7\,000\AA are used (``BE" technique, \citet{BE99}) for
this purpose. For this reason the comparison between the observed and simulated line ratios is 
a powerful method of validation for both the numerical code and the correct interpretation of observational
data.

\begin{figure}[!h]
\centering
\includegraphics[width=0.5\textwidth]{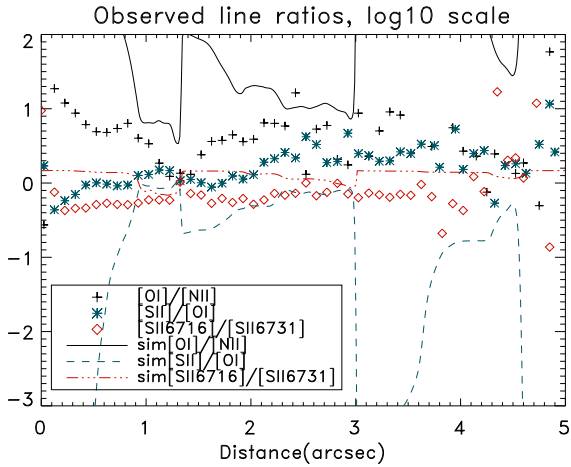}
\caption{\footnotesize Line ratios between the three doublets of [SII]6716+6731\AA, [OI]6300+6363\AA and [NII]6548+6583\AA,
                       log10 scale plot on the jet axis from simulation type $A$ without pre-ionization, and observations. }
\label{fig:em_lines_A2}
\end{figure}

In the previous 1D analyses we considered the emission of a single
shock at different times while propagating along the jet, 
instead we are now taking snapshots at given times of the whole length of the jet
and study the behaviour of the line ratios as a function of the longitudinal coordinate.
The high numerical resolution achieved thanks to the AMR technique
 allows us to follow not only the values in the
emission peaks, but also their evolution in the post-shock zone as the gas cools. 
We draw in Fig.\,\ref{fig:em_lines_A2} the results of the calculations,
without pre-ionization, of three line ratios of forbidden lines 
in comparison with the observed line ratios (symbols) for the first part of the redshifted 
jet from the RW Aurigae pair. We see that the values of
the calculated line ratios approach observations only for short distances after the shocks. 

\begin{figure}[!h]
\centering
\includegraphics[width=0.5\textwidth]{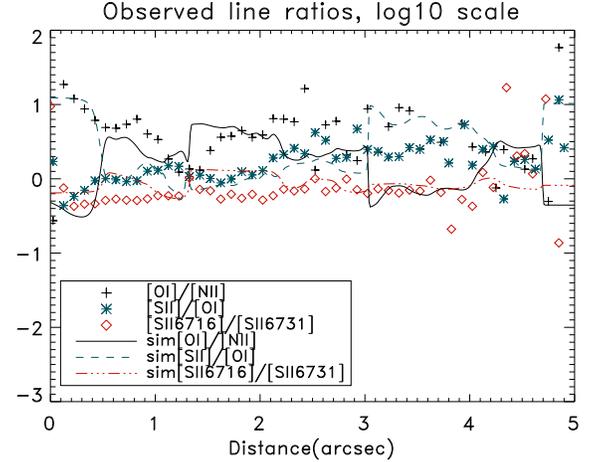}
\caption{\footnotesize Line ratios between the three doublets of [SII]6716+6731\AA, [OI]6300+6363\AA and [NII]6548+6583\AA,
                       log10 scale plot on the jet axis from simulation type $A$ with pre-ionization, and observations. }
\label{fig:em_lines_A}
\end{figure}

\begin{figure}[!h]
\centering
\includegraphics[width=0.5\textwidth]{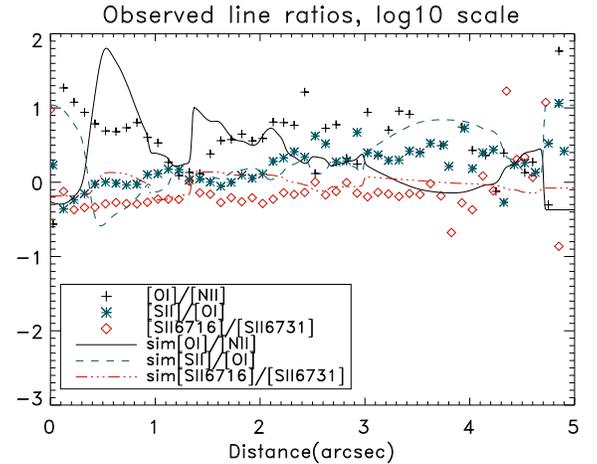}
\caption{\footnotesize Line ratios between the three doublets of [SII]6716+6731\AA, [OI]6300+6363\AA and [NII]6548+6583\AA,
                       log10 scale plot on the jet axis from simulation type $B$ with pre-ionization, and observations. }
\label{fig:em_lines_B}
\end{figure}

In Figs.\,\ref{fig:em_lines_A} and (Fig.\,\ref{fig:em_lines_B}) we show a simulation from 
the $A$ and $B$ sets, respectively, with pre-ionization included.
In both cases the behaviour of the calculated line ratios is much more consistent with 
observational data, the variations between knots remaining in the observed ranges.

\subsection{Position-Velocity diagrams}
%

In order to illustrate the distribution in velocities of the emitting material, the Position-Velocity (PV)
diagrams are widely used. A spectrum is generated for each pixel along the spectrograph slit, 
and the results are plotted in units of surface brightness at a certain wavelength on a 
Position-Velocity map.

\begin{figure} [!h]
\begin{center}
\begin{tabular}{c}
 \resizebox{85mm}{!}{\includegraphics{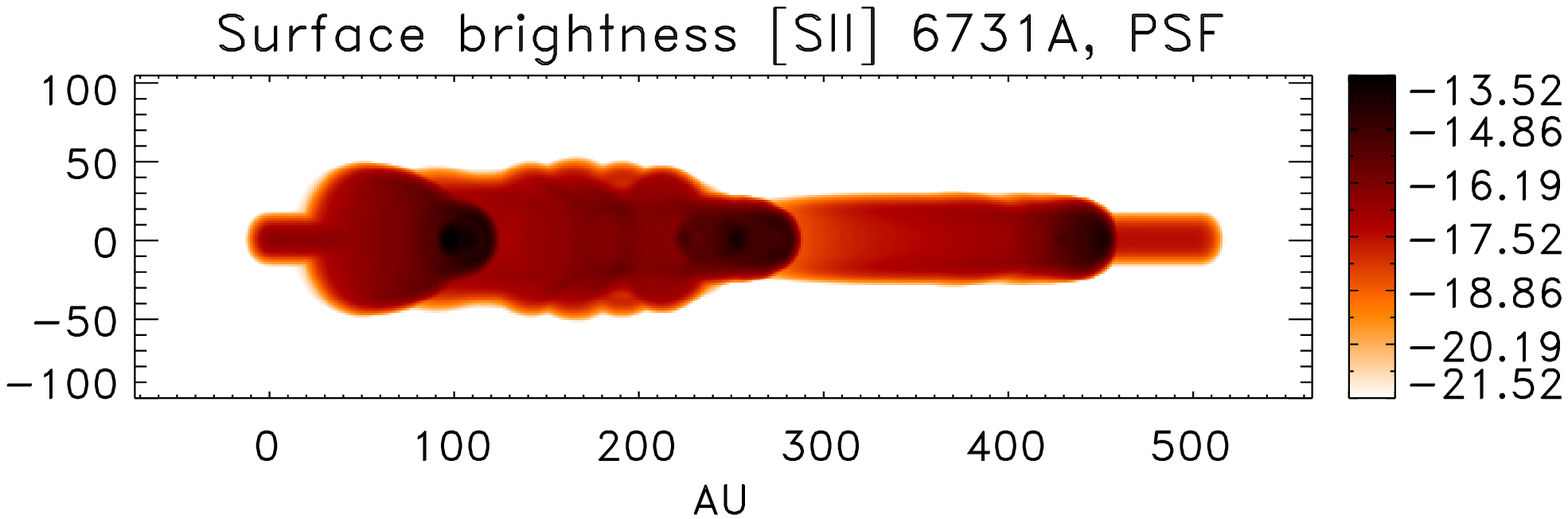}} \\
 \resizebox{85mm}{!}{\includegraphics{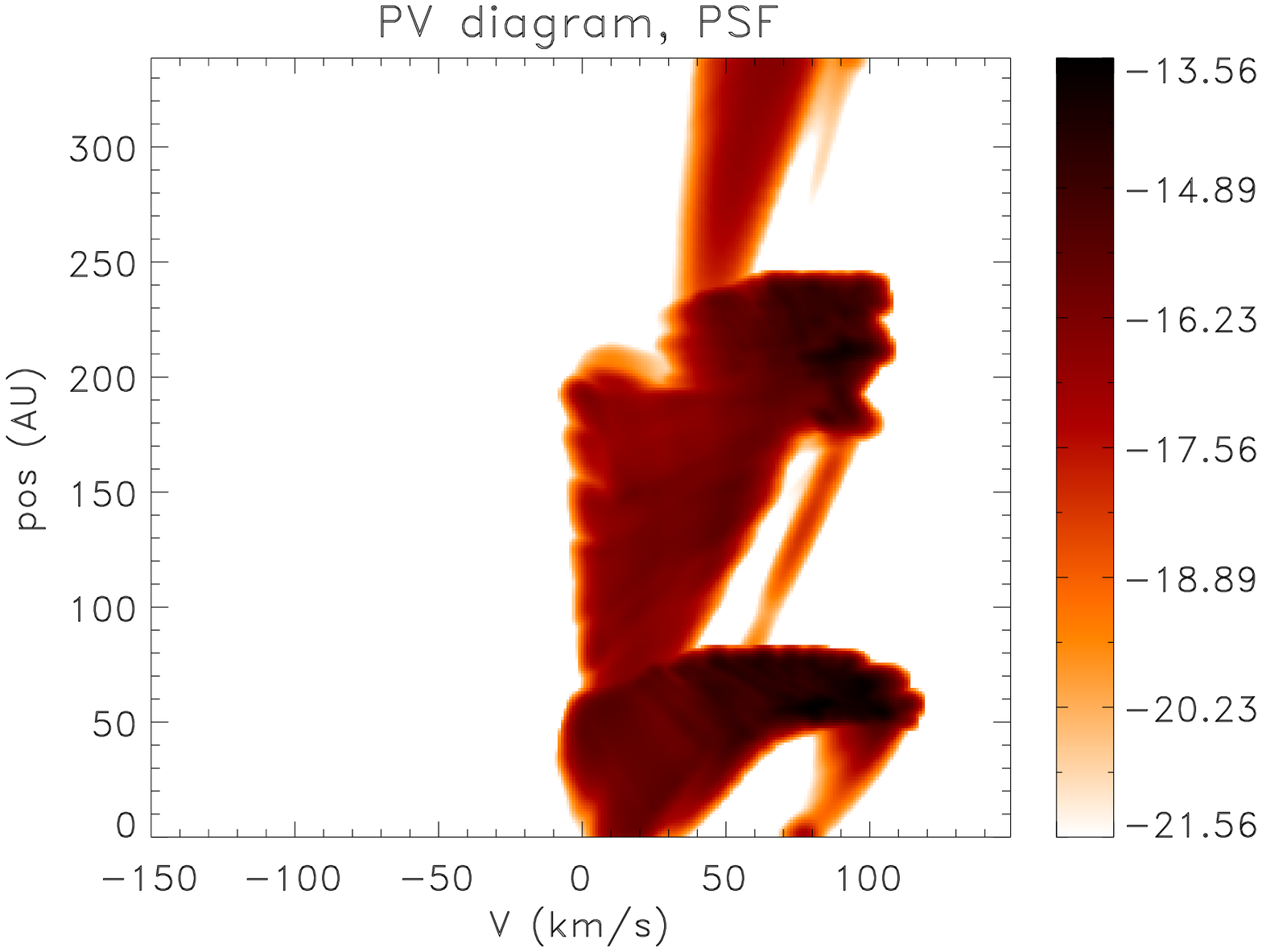}} 
\end{tabular}
\caption{Surface brightness (top panel) with the defined slit $0^{\prime\prime}.4$ 
wide (at the distance of Taurus), 
         and position-velocity diagram (bottom) for [SII] 6731\AA.}
\label{fig:pv_diag}
\end{center}
\end{figure}

If Fig.\,\ref{fig:pv_diag}, the output from the PLUTO post-processing routines is shown.
The top panel is a surface brightness map in one of the lines of the [SII] doublet, with
the user-defined slit from where the data for the PV-diagram will be taken. The bottom 
panel displays the resulting PV diagram, in units of surface brightness. The distribution
of brightness is concentrated to the right half of the image, corresponding to positive
velocities, due to the declination angle between the jet axis and the line of sight. 
The enhanced emission knots can be clearly seen in the PV diagram, concentrating around 
radial velocities of 90\,km\,s$^{-1}$. The inter-knot jet material is distributed in a range 
of velocities between -10 and +70\,km\,s$^{-1}$. 

 The PV diagrams are a powerful tool in the modern study of the structure of stellar jets, 
providing more accurate information on the velocity distribution of the emitting material. 
By the differences in the radial velocity and asymmetries between opposite parts of the jet,
their rotation (predicted by models) can be inferred \citep{CA07}. Consequently, as both 
the spatial and spectral resolutions of observational data increased, these diagrams were
geenrated also from the jet models, in order to be compared to the ones derived from 
observations \citep{CA06,SR07}. Arrays of models were devised \citep{KVR06}. 

An interesting study underway, where PV diagrams from multiple slits will be employed, 
focuses on DG-Tau and RW Aurigae, in the search for rotation signatures.

\section{Conclusions and summary}
%
%
%

Starting from numerical MHD simulations that include ionization network and detailed radiative cooling, 
we have obtained synthetic emission maps of surface brightness at various wavelengths relevant for 
observations of HH microjets. The comparison with observations was not limited to surface brightness
(along the jet, integrated in velocity), we have also tried to match the observed line ratios 
for different values of the simulation parameters.

We have shown the crucial role assumed by the pre-existing ionization in the jet medium, 
prior to the passage of the shock wave,
for the line emission properties of the corresponding ``knot". We believe that
pre-ionization will be a key ingredient in future work. 
This relatively high ionization fraction is likely to come from the X-ray
photoionization of the atoms at the jet base, being advected away with the flow 
conserving its value because of the low the recombination rate.
 The pre-ionization increases the number of free electrons in the gas and
speeds up the processes of ionization and excitation at the passage of the shock wave.

Among the simulations performed during this work, the $B$ and $D$ sets, that include a toroidal magnetic
field, rotation of the jet and pre-ionization, seem to compare well with observations. 
Future analyses will address the problem of the contrast between the knots and intra-knots brightness,
that remains higher than observed, for performing simulations aiming to reproduce in greater
detail the emission features of particular objects, with the goal to constrain the
jet physical parameters and better understand the physical mechanisms at work. Moreover,
a challenging but potentially insightful investigation will be the 3D case, that could address 
the shock misalignment.

\begin{acknowledgements} 
We are grateful to Prof. P. Hartigan (Rice University) 
and Prof. A. Glassgold (University of California) for insightful discussion.
OT was supported, in the Romanian PNII framework, by contract CNCSIS-RP no. 4/1.07.2009. 
The computational simulations were partly performed at CINECA Bologna, under the HPC-Europa2
project (project number: 228398) with the support of the European Commission -- Capacities
Area -- Research Infrastructures. Part of the simulations were performed using the computational 
resources of the CASPUR Supercomputing consortium.

\end{acknowledgements} 

\appendix
\section{Radial Equilibrium Solution}
\label{sec:equil}
%

The equilibrium solution is constructed by considering 
the radial force balance between pressure, magnetic and
centrifugal forces under the assumption $v_r=B_r=0$.
The equilibrium condition is expressed through the steady-state 
$r$-component of the momentum equation, which reads
\begin{equation}\label{eq:rad_equil}
  \frac{dp}{dr} = \frac{\rho v_\phi^2}{r} - \frac{1}{2}\left[
      \frac{1}{r^2}\frac{d(rB_\phi)^2}{dr}
      + \frac{dB_z^2}{dr}\right] \,.
\end{equation}
 In the present context, we will ignore the effect of a poloidal field component
and simply consider cases with $B_z=0$.
Density and longitudinal velocity profiles can be chosen to smoothly 
match their ambient values for $r>R_j$ while the azimuthal component
of magnetic field is prescribed by
\begin{equation}
B_\phi(r) = -\frac{B_m}{r}\sqrt{1 - \exp\left[-(r/a)^4\right]} \,,
\end{equation}
where $a=0.9$ is the magnetization radius and $r_j$ is the jet radius.
This choice guarantees that at large radii the field becomes essentially 
force-free whereas close to the axis the electric current $J_z\approx -2B_m r_j/a^2$
is approximately constant.
A convenient profile for the azimuthal velocity is
\begin{equation}
  v_\phi(r) = \alpha\frac{rr_j}{a^2}\sqrt{\frac{2\exp\left[-(r/a)^4\right]}
                                         {\rho}}\,,
\end{equation}
where the constant $\alpha$ sets the amount of rotation and the relative
importance of the centrifugal to the Lorentz force.
With these assumptions Eq. (\ref{eq:rad_equil}) can be integrated 
giving
\begin{equation}\label{eq:pequil}
  p(r) = p_j + \frac{1}{2}\frac{(\alpha^2-B_m^2)\sqrt{\pi}
         \,\textrm{erf}(r^2/a^2)}{(a/r_j)^2}\,,
\end{equation}
where $p_j$ is the jet pressure on the axis.
Clearly, when $\alpha>B_m$, the gas pressure increases monotonically with
$r$ while the opposite is true for $\alpha < B_m$. 
The condition $\alpha=B_m$ yields exact balance between
rotations and magnetic forces.

The actual value of $\alpha$ can be expressed in terms of the maximum 
rotation velocity $v_{\phi}^{\max}$ which, in the limit of constant density,
becomes
\begin{equation}
  \alpha \approx v^{\max}_{\phi}\left(\frac{e}{2}\right)^{1/4}\frac{a}{r_j}
         \sqrt{\rho_j}
\end{equation}
Finally, in order to specify the magnetic field strength $B_m$, we note
that, by assigning the equilibrium ambient temperature 
$T_a=p_a\mu_am_a/(\rho_ak_B)$ (where $\rho_a$ is the ambient density), 
Eq (\ref{eq:pequil}) may be solved 
for the magnetic field strength $B_m$ giving
\begin{equation}\label{eq:Bm}
  B_m^2 = \alpha^2 + 
          \frac{2k_B}{\sqrt{\pi}m_a}\left(\frac{a}{r_j}\right)^2
          \rho_j\left(\frac{T_j}{\mu_j} - \frac{T_a}{\eta\mu_a}\right)
\end{equation}
where $k_B$ is the Boltzmann constant, $m_a$ is the atomic mass unit, 
$\rho_j$ is the jet density, $\eta = \rho_j/\rho_a$ is the jet to ambient 
density contrast, $\mu_j$ and $\mu_a$ are the mean molecular 
weights in the jet and in the ambient medium, respectively.
Eq (\ref{eq:Bm}) immediately shows that, for $T_a < T_j$, the magnetic field 
has a lower threshold value and its strength always increases with rotation.

{}

\end{document}